\documentclass[twocolumn, tighten]{aastex701}

\usepackage{graphicx}
\usepackage{natbib}
\usepackage{url}
\newcommand\Msun{$\mathrm{M_\odot}$}
\newcommand\Teff{$T_{\rm eff}$}
\newcommand\vsini{$v \sin i$}
\newcommand\kms{$\rm{km\;s^{-1}}$}

\begin{document}

\title{Blue Straggler Stars in Old Open Clusters and the Kraft Break}
\correspondingauthor{Evan Linck}
\author[0009-0006-7474-7889]{Evan Linck}
\affiliation{Department of Astronomy, University of Wisconsin-Madison, 475 N. Charter St., Madison, WI 53706, USA}
\email{elinck@wisc.edu}

\author[0000-0002-7130-2757]{Robert D. Mathieu}
\affiliation{Department of Astronomy, University of Wisconsin-Madison, 475 N. Charter St., Madison, WI 53706, USA}
\email{mathieu@astro.wisc.edu}

\received{2026 April 17}
\revised{2026 May 12}
\accepted{2026 May 14}

\begin{abstract}
We measure the projected rotational velocities (\vsini) of the solar-like blue straggler stars (BSSs) in the old ($\geq4$ Gyr) open clusters M67, NGC 188, and NGC 6791. We find that the BSS rotation distribution shows a Kraft break similar to that found in the field. The main-sequence progenitors of these BSSs were cooler than the Kraft break and have spun down by their age. The binary interactions that create BSSs are expected to spin up these progenitors, so current BSS rotation rates are due to transformation and any subsequent spin-down. We observe that BSSs hotter than the Kraft break are rapidly rotating, showing that binary evolution spins up these, and likely all, BSSs to initial rotational periods below two days---still below critical velocity. BSSs below the Kraft break currently have slow rotation rates, and those within the Kraft break have a mixture of rotation rates suggesting rotational transition. This dependence of rotation on effective temperature indicates that BSS envelopes behave like those of single stars, becoming convective and generating magnetic fields at the same temperatures. For globular cluster BSSs with [Fe/H]$\sim-1.5$, we find evidence of a BSS rotation transition region that is 100--250 K hotter than at solar metallicity. We find the \vsini\ distributions of BSSs in open clusters have similar characteristics to both high- and low-density globular clusters, indicating the density of environment is not the only factor that can determine rotational distributions. We suggest that velocity dispersion plays an important role.

\end{abstract}


\section{Introduction} \label{sec:intro}

Blue straggler stars (BSSs) in star clusters are bluer, and often brighter, than the main-sequence turn-off (MSTO) of their host cluster. Their position on a color-magnitude diagram (CMD) indicates that they can be many tenths of a solar mass more massive than other cluster stars (\citealp[e.g.,][]{leinerCensusBlueStragglers2021,jadhavBlueStragglerStars2021,mathieuBlueStragglersFriends2025}; \citealt[submitted][]{linckDistributionBlueStragglerarxiv2026}). BSSs form after gaining mass through an interaction within binary stars via mass transfer, merger, and collisions \citep{mccreaExtendedMainSequenceStellar1964, andronovMergersClosePrimordial2006, leonardStellarCollisionsGlobular1989}. Each of these formation pathways is expected to spin-up the mass-gaining star (\cite{sunStellarSpinPostMass2024, schneiderStellarMergersOrigin2019, sillsHighresolutionSimulationsStellar2002}). In the mass-transfer scenario, a major challenge is that so much angular momentum is predicted to be transferred by even only a few hundredths of a solar mass of material that BSSs will be spun-up to critical velocity, at which point they stop accreting material \citep{packetSpinupMassAccreting1981, Matrozis+2017, sunStellarSpinPostMass2024}. 

Rotation is a critical diagnostic for many stellar processes. For stars with convective envelopes---those within and cooler than the Kraft break \citep{kraftStudiesStellarRotation1967}, magnetic braking provides a mechanism to spin a star down \citep{gossageMagneticBrakingMESA2023}, provided they have convective envelopes that yield efficient magnetic braking. The efficiency of magnetic braking has a complex relationship between the rotation rate of the star and the convective turnover time---which in turn is a function of the depth of the convective zone and has an empirical dependence on [Fe/H], \Teff, log \textit{g}, stellar evolution phase, and color \citep{bonannoAsteroseismicCalibrationRossby2025}. Whether a star has a convective envelope is set by the opacity of the outer layers of a star due to ionization zones, which is metallicity and temperature dependent, and the steepness of the temperature gradient needed to transport energy through the envelope. Roughly, solar-metallicity stars less massive than 1.35 \Msun\ will develop large enough convective envelopes for spin-down over the course of several hundred Myr to a few Gyr. Magnetic braking enables gyrochronology, which links low-mass-star rotation rates with their ages \citep{barnesRotationalEvolutionSolar2003, angusPreciseStellarAges2019, boumaEmpiricalLimitsGyrochronology2023}.

BSSs have been found to be rapidly rotating in a wide variety of environments, including the field \citep[e.g.,][]{prestonWhatAreThese2000,carneyMetalpoorFieldBlue2005}, open clusters \citep[e.g.,][]{leinerObservationsSpindownPostmasstransfer2018, linckWIYNOpenCluster2024}, and globular clusters \citep[e.g.,][]{ferraroFastRotatingBlue2023}. Here we study the distribution of BSS rotation rates in the open clusters M67, NGC 188, and NGC 6791 as a function of effective temperature (\Teff).

This work is the companion paper to \citet[submitted][hereinafter Paper 1]{linckDistributionBlueStragglerarxiv2026}, in which we investigated the stellar properties of BSSs in six old open clusters ($>$ 4 Gyr), including the three here, using their CMD locations. Here, we build on those BSSs and their properties. In Section~\ref{sec:bss_sample}, we first measure the projected rotational velocities (\vsini) of the BSSs. In Section~\ref{sec:OC_rot}, we examine the relationships of both \vsini\ and rotation periods with \Teff\ and discuss the implications of the results for angular momentum transfer during the interaction, BSS stellar structure, and gyrochronology. In Section~\ref{sec:gc_rotation}, we compare our open cluster findings to similar studies in globular clusters and explore the metallicity, density, and velocity dispersion dependencies of the \vsini\ distribution. Finally, we summarize our findings in Section~\ref{sec:summary}.

\section{BSS Sample and Measurements}\label{sec:bss_sample}

We examined the \vsini\ of BSSs in three old open clusters with MSTO masses $<1.3$\;\Msun\ that are well-studied by the WIYN Open Cluster Study \citep[WOCS; ][]{mathieuWIYNOpenCluster2000}: M67 (4.1 Gyr), NGC 188 (6.6 Gyr), and NGC 6791 (8.6 Gyr). Notably, main-sequence stars in these old clusters are both cooler than the Kraft break, meaning they have an efficient mechanism for spin-down, and old enough to have spun down \citep{boumaEmpiricalLimitsGyrochronology2023}. Given this spin-down of the BSS progenitors in these clusters, the rotation rates of present-day BSSs should be due to the binary interactions that created them and any subsequent spin-down.

This work uses the membership lists, isochrone fits, and BSSs identified and characterized (including \Teff, radius, and errors) in Paper 1. BSS and other cluster member star properties were derived using Gaia DR3 G, BP, RP photometry \citep[][]{GaiaDataRelease2023} and single-star MIST models \citep[Version 1.2,][]{dotterMESAISOCHRONESLAR2016, choiMESAISOCHRONESSTELLAR2016, paxtonMODULESEXPERIMENTSSTELLAR2011, paxtonMODULESEXPERIMENTSSTELLAR2013,paxtonMODULESEXPERIMENTSSTELLAR2015}. Figure~\ref{fig:cmd} shows the CMDs of each cluster, the \vsini\ of member stars, and the BSSs of each cluster. See Appendix~\ref{app:measurements} for BSS IDs, properties, and \vsini\ measurements. 

\begin{figure*}
    \centering
    \includegraphics[width=\linewidth]{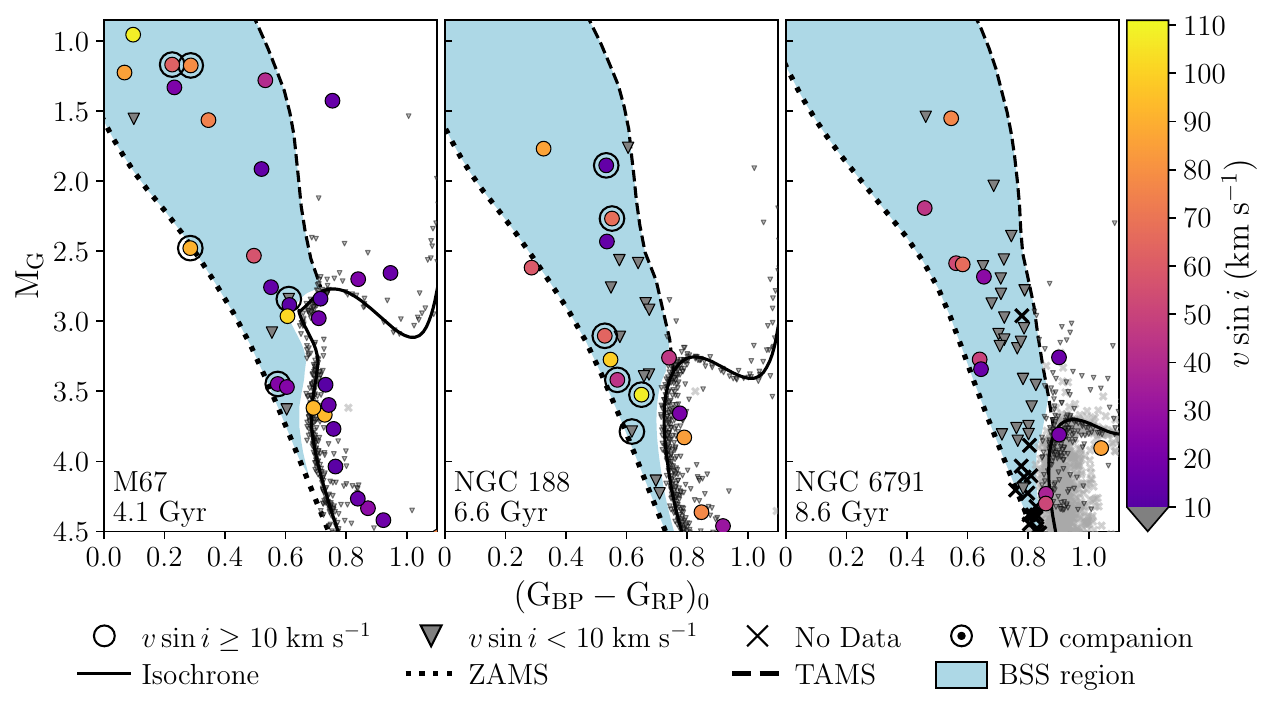}
    \caption{The BSS and upper main-sequence regions of the M67, NGC 188, and NGC 6791 CMDs. The BSS regions---bounded by the isochrone (solid line), ZAMS (dotted line), and TAMS (dashed line) found in Paper 1---are denoted by the blue areas. Stars with \vsini\;measurements above our floor of 10 km s$^{-1}$ are marked with color-coded circles; stars with measured \vsini\;below our floor are marked with downward triangles; stars without \vsini\;measurements are marked with black Xs. Non-BSSs that are slowly rotating or are lacking data are shown at a smaller scale for visual clarity of the main sequence. Stars with a ring around the central point have a known WD companion.}
    \label{fig:cmd}
\end{figure*}

The WOCS radial-velocity survey has made extensive time-series spectroscopic measurements of all stars on the upper main sequence and brighter for M67 and NGC 188 \citep{gellerStellarRadialVelocities2021, narayanWIYNOpenCluster2026} and most stars from the MSTO and brighter for NGC 6791 \citep{tofflemireWIYNOPENCLUSTER2014}. These measurements were made with the WIYN 3.5m Hydra Multi-Object Spectrograph (R$\sim20,000$).\footnote{The WIYN 3.5m Observatory is a joint facility of the University of Wisconsin–Madison, Indiana University, NSF’s NOIRLab, the Pennsylvania State University, and Princeton University.}

Following the procedure of \cite{rhodeRotationalVelocitiesRadii2001}, \vsini\ is measured through the \textsf{IRAF} task \textsf{fxcor} \citep{fitzpatrickIRAFRadialVelocity1993}, which calculates the cross-correlation function (CCF) between observed spectra and an observed solar template spectrum. To calibrate the relationship between \vsini\ and CCF full-width half-maximum (FWHM), we used artificially spun-up solar spectra, made by convolving our observed solar template with rotational profiles at specific \vsini\ up to \vsini\;$=150\rm{\;km\;s^{-1}}$\citep{gellerWIYNOPENCLUSTER2010}. These spun-up templates were then cross-correlated with the observed template to define the relationship between \vsini\ and CCF FWHM. Above \vsini\;$=10\rm{\;km\;s^{-1}}$, the FWHM strongly correlates with \vsini. However, the relationship flattens below \vsini\;$=10\rm{\;km\;s^{-1}}$ due to the spectral resolution limit; we take this as our measurement floor. 

For stars with multiple spectra (almost every BSSs in NGC 188 and M67 have $>10$ observations and those in NGC 6791 that we have observed have at least 3), we take the median \vsini. FWHM measurements are consistent across these multiple spectra (usually within 2\%), especially for stars with \vsini\;$<80\rm{\;km\;s^{-1}}$. We estimate errors on \vsini\ to be $<2\rm{\;km\;s^{-1}}$ for these stars, although errors rise rapidly above this to $\sim10\rm{\;km\;s^{-1}}$ for \vsini\;$=120\rm{\;km\;s^{-1}}$. 

For this work, we adopt the definition of fast rotating stars of \cite{ferraroFastRotatingBlue2023}: \vsini\;$\geq40$ \kms. Due to our measurement lower limit, we categorize slow rotators as those with \vsini\;$\leq10$ \kms. Intermediate rotators are between those limits. 

\cite{nineWIYNOpenCluster2024} has measured the \vsini\ of many of the BSSs in M67; our measurements agree to within $\sim5\rm{\;km\;s^{-1}}$ of theirs. APOGEE-2 has made \vsini\ measurements of 9 of our BSSs \citep[7 in M67 and 2 in NGC 188,][]{abdurroufSeventeenthDataRelease2022}; all but one of these measurements are within 5 $\rm{\;km\;s^{-1}}$ of ours (the remaining is 15 \kms\ higher than our measurement of 13 \kms). APOGEE-2 has measured the \vsini\ of WOCS 2011 in M67 to be 8.84 km s$^{-1}$, which we adopt for this study. Although there are some differences in the \vsini\ measured for these stars, each of these studies agrees that the same sets of stars are rapidly or slowly rotating. 

Tidal forces will impact the rotation of close binaries. We follow the method of \cite{leinerBlueLurkersHidden2019}, removing any stars with known orbital periods less than twice the empirical tidal synchronization limit of $\sim$30 days \citep{lurieTidalSynchronizationDifferential2017}: WOCS 4003 and 1007 in M67; WOCS 4230 and 5078 in NGC 188; and WOCS 54008 in NGC 6791. We further remove WOCS 2009 in M67 and WOCS 5885 in NGC 188 due to inability to measure photometric \Teff\ given multiple bright members in the systems. 

\section{BSS Rotation in Old Open Clusters}\label{sec:OC_rot}

We measure the numbers of fast, intermediate, and slow rotators in each cluster: M67, $7\;(44 \pm 17\%)$, $5\;(31 \pm 14\%)$, and $4\;(25 \pm 12\%)$; NGC 188, $5\;(26 \pm 12\%)$, $2\;(11 \pm 7\%)$ and $12\;(63 \pm 18\%)$; NGC 6791, $4\;(13 \pm 6\%)$, $2\;(6 \pm 5\%)$, and $25\;(81 \pm 16\%)$, respectively. We note that the fraction of fast rotators decreases with cluster age. 

\subsection{Temperature Distribution of \vsini}\label{subsec:teff_vsini}

\begin{figure*}
    \centering
    \includegraphics[width=\linewidth]{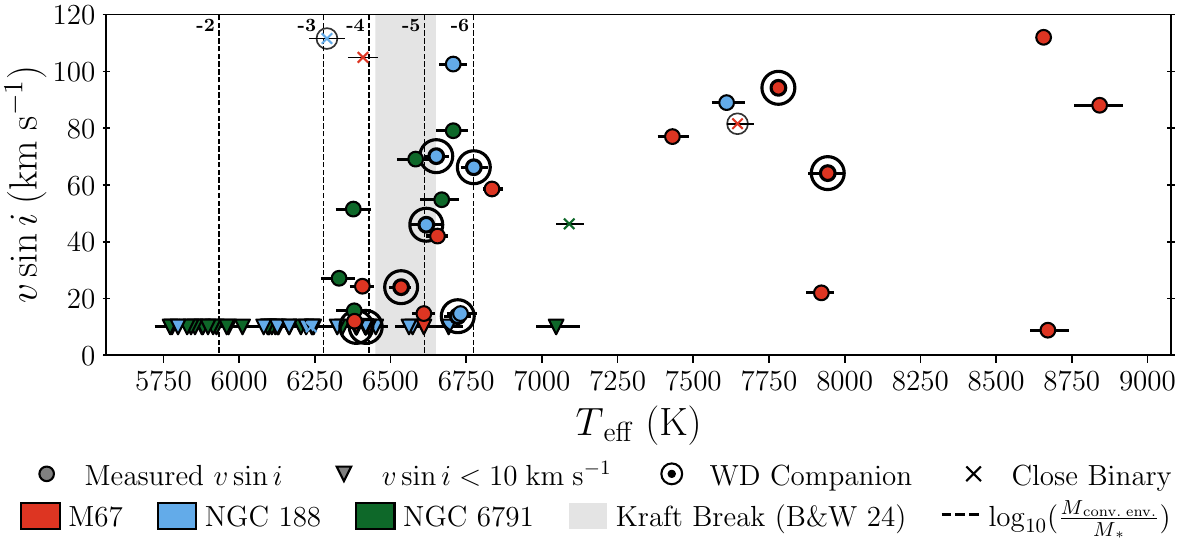}
    \caption{The BSS \vsini\;distribution by effective temperature, color-coded by cluster (M67: red, NGC 188: blue, NGC 6791: green). BSSs used in the analyses in this work with a measured \vsini\;are marked by circles; those with a \vsini\;below our floor of 10 km s$^{-1}$ are marked with triangles at their upper limit. BSSs that have been discarded due to an orbital period less than 40 days are marked with an X. BSSs without a \vsini\; measurement are not plotted. Finally, BSSs with a known WD companion have a ring around their central point. Three distinct populations are apparent: cool stars that are all slowly rotating, hot stars that are primarily rapid rotators, and a population in between with a mixture of rotation rates. This final group are stars with thin convective envelopes that will spin-down, but slower than cooler stars with thick convective envelopes. For reference, we plot a dashed line at the maximum \Teff\;at which the mass fraction of a convective envelope decreases by an order of magnitude among our MESA models. The region of a mixture of BSS rotation rates and thin convective envelopes aligns well with the empirical Kraft break region of \citet[plotted as the grey-shaded area from 6450--6650 K]{beyerKraftBreakSharply2024}, although is both cooler and hotter (roughly 6215--6750 K).}
    \label{fig:teff_vsini_OC}
\end{figure*}

In Figure~\ref{fig:cmd}, we observe that the fastest rotating stars are at bluer colors than slower rotating stars. At colors bluer than $\rm BP-RP = 0.6$, BSS \vsini\ are as high as 110 \kms. At colors redder than $\rm BP-RP = 0.6$, most BSS \vsini\ drop below our measurement floor. 

We plot \Teff\ versus \vsini\ in Figure~\ref{fig:teff_vsini_OC}. The \vsini\ distribution naturally divides into three distinct regions: below 6300 K, all of the BSSs have \vsini\;$\leq$ 10 \kms; between 6300 K and 6750 K, the BSSs show a mix of rotation rates; and above 6750 K all but two of the BSSs are measurably rotating. We also identify BSSs with known WD companions \citep{gosnellIMPLICATIONSFORMATIONBLUE2015, sindhuUVITOpenCluster2019,jadhavUVITOpenCluster2019, gosnellConstrainingMasstransferHistories2019,leinerObservationsSpindownPostmasstransfer2018, vernekarPhotometricVariabilityBlue2023, nineDetectionBLM672023, palDiscoveryBariumBlue2024}.

First, we examine the 11 hottest BSSs (\Teff\;$> 6750$\;K) without close companions. The 9 measurably rotating stars have an average \vsini\;of $75 \pm 8 \rm\; km \;s^{-1}$ ($\sigma = 24\;\rm km \;s^{-1}$). Assigning each star the median inclination angle ($i=60^\circ$) gives $\overline{v}_{\rm rot} = 87 \pm 9 \rm\; km \;s^{-1}$. The two stars that have \vsini\;$\leq$ 10 km s$^{-1}$ (NGC 6791: WOCS 4003; M67: WOCS 2011) are among the hottest and most massive stars in each cluster. Using a Monte Carlo simulation to estimate the true distribution of \vsini\ from our mean \vsini\ and error at $i=60^\circ$ as the mean $v_{\rm rot}$ of rotators and a random uniform distribution of the cosine of the inclination angle, we find $p = 7\%$ that 1 star of 11 has a \vsini\ $<10\;\rm km \;s^{-1}$, but $p = 0.2\%$ that 2 stars do. This means that in general, the \vsini\ distribution of the hottest BSSs are consistent with a population in which all BSSs are rapidly rotating, although an exception or two may exist. Examining the population below 6300 K shows that none are measurably rotating, let alone rapidly rotating. Finally, the population between 6300 and 6750 K shows a mix of rotation rates, with a trend to lower rotation rates at cooler temperatures. 

These divisions align closely with the well-known Kraft break. In the literature, the division between hot and cold stars (those with or without a convective envelope) is frequently given as \Teff\ $\sim6200$ K without citation. In a recent empirical study of rotation rates of nearby field stars, \cite{beyerKraftBreakSharply2024} found stars with \Teff\ between 6450 and 6650 K to have a mixture of rapidly rotating and slowly rotating stars and labeled this temperature domain as the Kraft break. We plot this region in Figure~\ref{fig:teff_vsini_OC} for reference and note that although it demarcates the area where BSS rotation rates are changing, there are BSS fast rotators that are somewhat cooler than this region and BSS slow rotators that are somewhat hotter than this region.

\cite{spaldingTidalErasureStellar2022} used a limit of the convective envelope accounting for $3\times10^{-3}$ of the total mass fraction of a star to identify the lower boundary of the Kraft break; we follow a similar approach here. In Paper 1, we used Modules for Experiments in Stellar Astrophysics \citep[MESA; version 24.03.01][]{paxtonMODULESEXPERIMENTSSTELLAR2011, paxtonMODULESEXPERIMENTSSTELLAR2013, paxtonMODULESEXPERIMENTSSTELLAR2015,paxtonModulesExperimentsStellar2018,paxtonModulesExperimentsStellar2019, jermynModulesExperimentsStellar2023} to model main-sequence solar-metallicity stars. We used MESA's predictive mixing algorithm with the Ledoux criterion. In Figure~\ref{fig:teff_vsini_OC}, we show the \Teff\ at which mass fractions of the convective envelopes of solar-metallicity main-sequence stars with masses between 1 and 1.5 \Msun\ decrease by orders of magnitude. At \Teff\;$\sim6750$ K, the convective envelopes rapidly decline with increasing temperature by several orders of magnitude and becomes negligible. 

The \Teff\ region of mixed BSS rotation rates maps very well to the decreasing mass fractions of convective envelopes. Among BSSs near the Kraft break, the coolest measurably rotating BSS has \Teff\;= 6330 K and the hottest slowly rotating BSS has \Teff\;= 6690 K. For specificity, we identify the BSS Kraft break as spanning \Teff\ of $6200-6750$ K and Gaia BP-RP of $0.66-0.52$, the \Teff\ domain where the mass fraction of the convective envelope falls from $10^{-3}$ to $10^{-6}$. This may generalize to the Kraft break for field stars. 

This in turn provides the explanation of the three rotation-rate populations among the BSSs. BSSs below the Kraft break have thick convective envelopes, providing an efficient mechanism for spin-down. BSSs that are above the Kraft break do not have an efficient mechanism to spin down and thus maintain their post-interaction rotation rate. BSSs within the Kraft break have thin convective envelopes that allow them to spin down, but do so slowly. This can be further seen across this region as the fastest rotators are at the hottest temperature, indicating that they have the thinnest convective envelopes that would take the longest to spin down.

In Paper 1, we found that the bluer BSSs formed within the last 1--2 Gyr. Redder BSSs had less restriction on their time as a BSS, but in M67 and NGC 188 generally had transformed in the last 1-4 Gyr, whereas most in NGC 6791 must have transformed within last 5 Gyr. The youngest age that these clusters could have been at the times when their current BSSs formed is about 2 Gyr, roughly the present age of NGC 6819 (2.5 Gyr), although most of the BSSs have formed more recently (including many within the last Gyr). Examining the MIST temperatures and rotation rates \citep{boumaEmpiricalLimitsGyrochronology2023,van-laneChronoFlowDatadrivenModel2025} of main-sequence stars in NGC 6819 and M67 shows that the progenitor main-sequence stars of most, if not all, of the BSSs would have low-enough \Teff\ to have convective envelopes and be old enough to have spun down to periods greater than 10 days by the time of the interaction. This means that the BSS rotation rates are due to the spin-up of the interaction and any subsequent spin-down due to magnetic braking.

There are several important implications of these observations. The first is that almost all of the BSSs that do not have an efficient spin-down mechanism are rapidly rotating, despite their progenitor accretors having previously spun down, indicating that BSSs are indeed spun-up during their interactions. Second, in order to exhibit the same \vsini--\Teff\ distribution of single stars, BSSs must develop convective envelopes at the same temperatures that single stars do, indicating that their surface structure is similar to a star undergoing normal single-star stellar evolution despite having accreted material. This finding matches that of \cite{leinerObservationsSpindownPostmasstransfer2018} in lower-mass interaction products with known WD companions---including several of the BSSs in NGC 188---that spun down following the gyrochronology relationships of low mass stars, which we explore in the next subsection. Third, the fraction of fast rotators BSSs tracks the fraction of BSSs in and above the Kraft break, which is a consequence of the MSTO mass (and thus whether progenitors would have spun down) and BSS masses of a given cluster age. Thus, old clusters have fewer fast rotators. 

\subsection{BSS Rotation Periods}\label{subsec:rot_per}

\begin{figure}
    \centering
    \includegraphics[width=\linewidth]{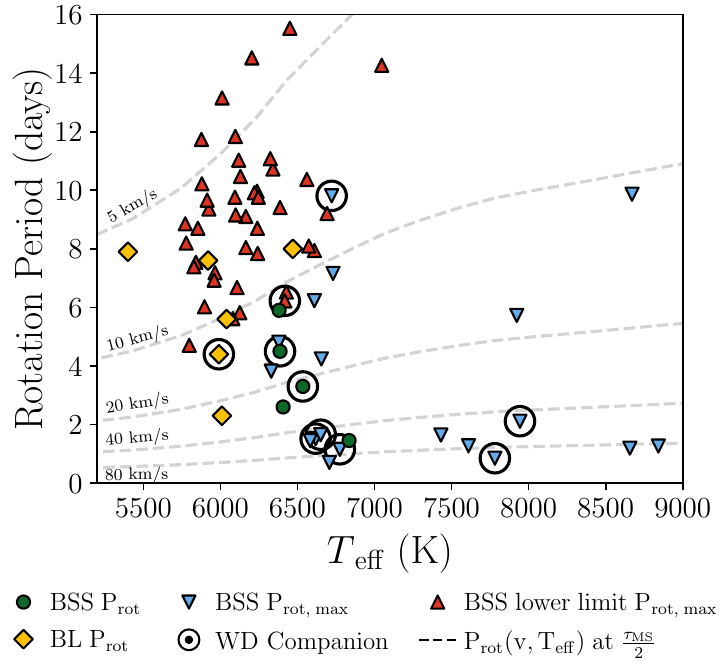}
    \caption{The rotation periods of the BSSs and BLs by \Teff. BSSs and BLs with known rotation periods are plotted with green circles and yellow diamonds, respectively. BSSs with measured \vsini\ have a maximum rotation period and are plotted with downward blue triangles. BSSs with \vsini\;$<10$~\kms\;have a known lower limit on maximum rotation period and are plotted with upward red triangles. Using MIST evolutionary tracks, we mapped the \Teff\ of main-sequence stars that are halfway through their main-sequence lifetimes to their radii at that time. We can then calculate the rotation period of these stars for a given equatorial velocity (i.e., \vsini\ at an inclination of $90^\circ$); we show these velocities as dashed lines. The radius of a star grows throughout its main-sequence lifetime, so these lines will shift lower for a star that is closer to the ZAMS and higher for a star closer to the TAMS.
}
    \label{fig:teff_P_OC}
\end{figure}

In the previous sub-section, we saw a striking relationship between \vsini\ and \Teff\ among the BSSs. In order to compare these results to other observational and theoretical studies \citep[e.g.,][]{leinerObservationsSpindownPostmasstransfer2018,nineDetectionBLM672023,sunStellarSpinPostMass2024}, rotational periods are needed. 

Direct measurements of rotational periods of BSSs in open clusters through photometric light curves can be challenging as the dense environments can pollute light curves with signals from neighboring stars (e.g., TESS pixels are 21 arcsec on a side). Further, the hottest BSSs in old open clusters are A- and early F-type stars, and less than one-third of A-type stars show rotational signatures (e.g., starspots) in their light curves \citep{balonaRotationalLightVariations2011}. Both M67 and NGC 6791 were observed by the Kepler spacecraft; several of the BSSs in these clusters have published rotation periods from these observations \citep{leinerBlueLurkersHidden2019, sanjayanVariableStarPopulation2022}. NGC 188 has been observed by TESS and light curves are available for some stars through differential image subtraction \citep{boumaClusterDifferenceImaging2019}, but to our knowledge no rotation periods of BSSs have been published. 

For most of the BSSs, we use \vsini\ values measured here and radii from Paper 1 to derive maximum rotation periods (due to $\sin i$). The \vsini-derived maximum periods are consistent with the rotation periods of the BSSs found through light curves. For stars with \vsini\;$\geq 10$ \kms, we can put a specific upper limit on the rotation period. For stars with \vsini\;$<10$ \kms, we can put a lower limit on the maximum period. We show the distribution of the BSS rotation periods with \Teff\;in Figure~\ref{fig:teff_P_OC} and mark stars that have hot WD companions. 

Blue lurkers (BLs) are mass-gaining stars embedded among the main sequences of clusters. These stars were first identified by \cite{leinerBlueLurkersHidden2019} in M67 due to their anomalously fast rotation periods (from light curves) despite being in long-period binaries, which indicates they were spun-up during a binary interaction. These stars are BSSs that did not gain sufficient mass to stand apart from other main-sequence stars in CMDs \citep{mathieuBlueStragglersFriends2025}. We also show them on Figure~\ref{fig:teff_P_OC}. (We note that none of the BLs of M67 have measured \vsini\ $>10$ \kms.)

The most rapidly-rotating BSSs at effective temperatures in and above the Kraft break have rotation periods below 2 days. None of these stars have equatorial velocities near critical velocities (350--450 \kms\ for these masses and radii). At most the rapid rotators are rotating at speeds about a quarter of critical velocity. 

It is of interest to compare the BSS rotation periods to empirical spin-down times \citep{boumaEmpiricalLimitsGyrochronology2023}. BSSs with $6000<T_{\rm{eff}}\lesssim 6200$ have a lower limit on maximum period greater than 6 days. Empirical rotation periods of stars in this temperature range increase quickly during the first 500 Myr from 1--2 days to about 4--6 days before the rate of change dramatically slows down. By 2.5 Gyr, stars in this temperature range have rotation periods of 7--10 days. Presuming initial rotation periods of $\sim1$ day as observed for the hotter BSSs, these cool BSSs with rotation periods of 7--10 days have transformation ages greater than 2 Gyr. These cool BSSs are low-mass BSSs in either of NGC 188 or NGC 6791 (Figure~\ref{fig:teff_vsini_OC}). As shown in Paper 1, many of the low-mass BSSs in NGC 188 do in fact have transformation ages permitting these spin-down timescales. Presumably this is also true in the even older NGC 6791.

An alternative approach is comparison of spin-down times with the transformation ages of the BSSs with known WD companions (all have cooling ages of $\lesssim600$ Myr). Unfortunately, all such BSSs except the BL with a hot WD \citep{leinerBlueLurkerWOCS2025} are above the temperatures at which spin-down rates have been modeled \citep[e.g., \Teff\;$<6200$ K,][]{boumaEmpiricalLimitsGyrochronology2023}. \cite{leinerBlueLurkerWOCS2025} found the WD companion of the BL to have a cooling age of 400 Myr. The BL rotation period of 4.4 days agrees well with the spin-down models of \cite{boumaEmpiricalLimitsGyrochronology2023}. Most of the BSSs with WD companions that are in the Kraft break have short periods ($\lesssim4$ days), which is consistent with their young transformation ages. However, two of these BSSs have longer (6--10 day) maximum rotation periods. Low inclination angles could explain both of these stars; light curves are needed to confirm these slow periods. 

Lastly, there are some slowly rotating BSSs (without known WD companions) in and above the Kraft break (lower limit on $P_{\rm{rot, max}}>10$ days and hotter than 6200 K). Inclination angle may explain some of these stars. These stars do have larger radii (all larger than $2\;R_\odot$) and most are near the ends of their main-sequence lifetimes (Paper 1 and Appendix~\ref{app:measurements}). The combination of increasing radii post-interaction and longer times as BSSs may have further spun them down. 

The findings in this section pose issues for the possibility of gyrochronology with BSSs in many clusters. In old open clusters, many of the BSSs are cool enough to have thin convective envelopes and spin down. \cite{sunStellarSpinPostMass2024} modeled post-interaction products of 1.2 \Msun\ and lower and found that they did spin down and explained the rotation distribution found by \cite{leinerObservationsSpindownPostmasstransfer2018}. However, in open clusters younger than 4 Gyr, the mass of a star at the MSTO is above 1.3 \Msun, meaning that most of the BSSs in these clusters are going to be hotter than the Kraft break. 

\section{Comparison to BSS Rotation Rates in Globular Clusters}\label{sec:gc_rotation}

\cite{ferraroFastRotatingBlue2023} compiled \vsini\ measurements of BSSs (1.1--1.5 \Msun) in eight globular clusters and found that fast rotating BSSs preferentially occurred in less dense globular clusters. The authors argue that the difference in BSS rotation rates is indicative of recent ($<1$ Gyr) BSS formation via mass transfer in the low-density clusters and past collisions (at least 1--2 Gyr ago) in high-density clusters (or that collision products have undefined very efficient spin-down mechanisms).

In this section we compare the rotation rate distribution with the \Teff\ of BSSs in open clusters to those of these globular clusters. The globular clusters included and their \vsini\;source papers (along with the fraction of rapidly rotating BSSs from \citealt{ferraroFastRotatingBlue2023}) are the low-density clusters M55 \citep[0.47,][]{billiRotationalVelocitiesBlue2024}, NGC 3201 \citep[0.28,][]{billiFastrotatingBlueStraggler2023}, $\omega$ Centauri \citep[0.41,][]{mucciarelliSPINNINGBLUESTRAGGLER2014}, and M4 \citep[0.40,][]{lovisiFastRotatingBlue2010}; and the dense clusters 47 Tucanae \citep[0.47,][]{ferraroDiscoveryCarbonOxygendepleted2006}, M30 \citep[0.06,][]{lovisiFLAMESXSHOOTERSPECTROSCOPY2013}, NGC 6752 \citep[0.0,][]{lovisiANOTHERBRICKUNDERSTANDING2013}, and NGC 6397 \citep[0.13,][]{lovisiChemicalKinematicalProperties2012}.\footnote{\vsini\;information was available at \url{http://www.cosmic-lab.eu/Cosmic-Lab/BSS_rotation_catalogs.html}} We also include the \vsini\ data from the high density globular cluster NGC 1851 \citep{billiFastRotatingBlue2026}, which has a reported floor on \vsini~of 15 \kms. Several stars in $\omega$ Centauri, 47 Tucanae, and M30 have been identified as contact binaries \citep{mucciarelliSPINNINGBLUESTRAGGLER2014, ferraroFastRotatingBlue2023}. As with the open clusters above, we remove these from our analyses.

\subsection{The Influence of Metallicity on BSS Rotation Distributions}

Metallicity changes the opacity profile of a star \citep{amardFirstGridsLowmass2019, amardImpactMetallicityEvolution2020}. \cite{spaldingTidalErasureStellar2022} found evidence for a metallicity-dependent ($-0.3<\rm{[Fe/H]}<0.3$) Kraft break that increased in \Teff\ at lower metallicity among the stellar obliquities of systems with hot Jupiters. \cite{amardEvidenceMetallicitydependentSpin2020} showed that the rotation rates of high-metallicity stars were slower than low-metallicity stars in the Kepler field. Notably, \cite{beyerKraftBreakSharply2024} did not find evidence of a metallicity-dependent Kraft break above and below the median metallicity of their field-star sample ([Fe/H] $ = -0.43$), but they note their sample size was small. At the much lower metallicity of [Fe/H] $ = -1.58$, \cite{billiFastrotatingBlueStraggler2023} noticed a trend between color and BSS rotation rate in the globular cluster NGC 3201, stating that this could be evidence of more massive BSSs having formed more recently or could be due to reduced convective envelopes at higher temperatures. Here, we examine the rotation rates as a function of temperature for the BSSs in four metal-poor ($-1.2<\rm{[Fe/H]}<-1.8$) low-density globular clusters. 

\begin{figure}
    \centering
    \includegraphics[width=\linewidth]{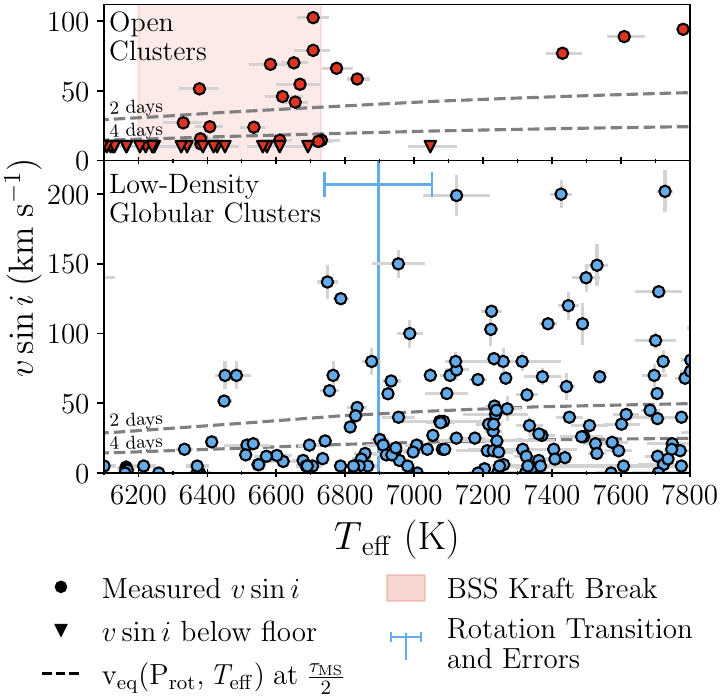}
    \caption{We show the \Teff--\vsini\ distribution of BSSs in open clusters (upper panel, in red, from Figure~\ref{fig:teff_vsini_OC}) and low-density globular clusters (lower panel, in blue). Like in Figure~\ref{fig:teff_P_OC}, using MIST evolutionary tracks, we calculated the equatorial velocity for stars that are half-way through their main-sequence lifetimes at a specific \Teff\ with rotational periods of 2 and 4 days, and plot these with dashed lines. We further show the BSS Kraft break at solar metallicity (red area) in the upper panel and the best-fit \Teff\ with errors (blue line) that demarcates a transition of the globular cluster BSS rotation distribution in the lower panel.}
    \label{fig:teff_vsini_GC}
\end{figure}

We derived photometric temperatures of each BSS with a \vsini\ measurement in M55, NGC 3201, M4, and $\omega$ Centauri from the photometry of \cite{stetsonHomogeneousPhotometryVII2019} and distances of \cite{baumgardtAccurateDistancesGalactic2021} as listed in \cite{hilkerGalacticGlobularClusters2019}. Stars were differentially dereddened using the average E(B-V) measurements listed in \citet[from which we also drew {[Fe/H]}]{harrisCatalogParametersGlobular1996} and differential reddening maps of \cite{pancinoDifferentialReddening482024}\footnote{Although the central region of $\omega$ Cen as having less reliable differential reddening data, we did include the 43\% of BSSs in this region.} and the bandpass reddening relations of \cite{schlaflyMeasuringReddeningSloan2011}. We used MIST to estimate a color-magnitude-temperature relationship. For each cluster (i.e, for a given [Fe/H] and distance), MIST provides a tight relationship between $\rm V-I$ color, V magnitude, and \Teff. To propagate photometric errors for each star, we implemented a Monte Carlo simulation that sampled from normal distributions of each star's photometry and photometric error. We note that there are no reported errors on metallicity or average E(B-V) in \cite{harrisCatalogParametersGlobular1996}, which would also contribute to uncertainties on \Teff.

We plot \Teff\ and \vsini\ of each BSS in the bottom panel of Figure~\ref{fig:teff_vsini_GC}. Most metal-poor BSSs below \Teff\;$\sim6750$ K are spun down. Above \Teff\;$\sim6750$ K, the metal-poor BSSs show a broad distribution in \vsini\ from a few \kms\ to 200 \kms, including a large increase in the frequency of BSSs with \vsini\ $<50$ \kms ($ P_{\rm{rot}}> 2$ days). For comparison, we re-plot the open cluster BSSs of Figure~\ref{fig:teff_vsini_OC} in the top panel. The rotation distribution for cooler globular cluster stars is similar to that of the open clusters and reminiscent of the Kraft break, but shifted to higher temperatures. 

We created a bootstrap method to search for a \Teff\ demarcating where a shift in rotational distribution occurs. We fit a step function to the data from 6100--7800 K using a quantile regression of the median (i.e., the y-value of each step was the median in that interval) as implemented in \textsf{statsmodel} \citep{Skipper_statsmodels_Econometric_and_2010}. For each of 500 draws, we found the \Teff\ of the discontinuity that best reproduced the data and the median of the distributions before and after the step. We found the rotation distribution changes at $6896\pm 156$ K, with the median \vsini\ shifting from $11.5 \pm 4.3$ \kms\ below to $33.7 \pm 4.8$ \kms\ above the break. 

This transition temperature is 100--250 K hotter at [Fe/H]\;$\sim-1.5$ than it is at solar-metallicity. We hypothesize this difference is due to metal-poor BSSs being able to undergo efficient magnetic braking at hotter \Teff. At this metallicity, from MIST models, the temperature range of the transition corresponds to stars of masses of 1.1 to 1.2 \Msun. Because these stars are lower mass, they may be able to maintain some convective envelopes at these hotter \Teff\ than solar-metallicity stars. We suggest that this phenomena is akin to the Kraft break of metal-rich stars. 

High-density globular clusters may primarily create BSSs through collisions, but mass transfer does still happen as evidenced by W UMa stars \citep{ferraroFastRotatingBlue2023}. Collision products may have different stellar structures \citep[e.g., radiative envelopes at low mass,][]{sillsHighresolutionSimulationsStellar2002}, and so may not spin down in the same manner \citep{sillsBlueStragglersStellar2005} as mass transfer stars, and thus not be impacted by the location of the Kraft break. Applying the same temperature-fitting procedure as above to the high-density globular clusters,\footnote{NGC 6397 is not in \cite{stetsonHomogeneousPhotometryVII2019}; we instead used the photometry listed in \cite{lovisiChemicalKinematicalProperties2012} and E(B-V) of \cite{harrisCatalogParametersGlobular1996}.} we find that the few fast rotators are hotter than 7000 K, which may provide further evidence that any mass-transfer products cooler than that limit are able to spin down whereas those above cannot. For example, the \Teff\ distribution of NGC 1851 mirrors that of the open clusters, with all of the stars cooler than 7000 K slowly rotating and two-thirds of those hotter than 7000 K rotating faster than 40 \kms. 

Additionally, we note that 80\% of the BSSs with \vsini\ measurements in 47\,Tuc and 72\% of those in NGC 6752 are cooler than 7000 K, which could contribute to the large fraction of non-rapid rotators found by \cite{ferraroFastRotatingBlue2023} if many of these stars have an efficient spin-down mechanisms.  

Two observational features noted above require further observations and modeling of the physics to understand the driving factors of the rotation distribution in metal-poor globular clusters. First, how does the interplay between metallicity, mass, and magnetic braking impact the \Teff\ of the Kraft break. For example, a metallicity dependence may have further evidence in Figure~\ref{fig:teff_vsini_OC}: the two coolest rapidly rotating BSSs are both in the metal-rich NGC 6791 ($\rm{[Fe/H]}\simeq0.35$), suggesting it may have a cooler Kraft break than M67 and NGC 188 (both of $\rm{[Fe/H]}\simeq0.0$). Second, the many high-\Teff\ low-\vsini\ BSSs that exist in the high-density globular clusters but do not exist in the old open clusters need to be explained. These BSSs may indicate that some difference in physical processes---such as mergers and their subsequent spin-down---is occurring between these environments.

\subsection{The Influence of Density on BSS Rotation Distributions}

\begin{figure}
    \centering
    \includegraphics[width=\linewidth]{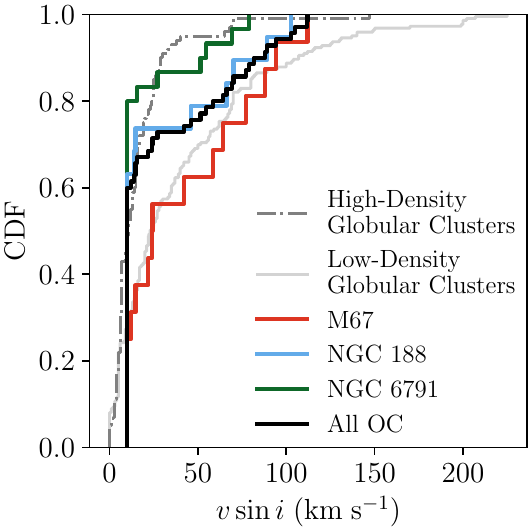}
    \caption{Comparison of the rotational velocity distributions of open and globular clusters using the empirical CDFs of three open clusters of different ages (M67: red, NGC 188: blue, NGC 6791: green, all: black) and the CDFs of (high-density  and low-density globular clusters (Figure 2 of \cite{ferraroFastRotatingBlue2023}).}
    \label{fig:vsini_radius_cdf}
\end{figure}

\cite{hilkerGalacticGlobularClusters2019} measured the central densities of these nine globular clusters, finding the low-density clusters to have $\log_{10} \rho_c\rm{\;(M_\odot\;pc^{-3})}$ between 2.54 and 3.99 and the high-density clusters between 4.72 and 6.74. In comparison, the best-fit SPES models of \cite{alvarez-baenaLongevityOldestOpen2024} give central densities for NGC 188 and M67 of $16-18 \rm{\;M_\odot\;pc^{-3}}$ and for NGC 6791 of $24\;\rm{M_\odot\;pc^{-3}}$ ($1.2<\log_{10} \rho_c\rm{\;(M_\odot\;pc^{-3})}<1.4$), all an order-of-magnitude less dense than any of the globular clusters.

In Figure~\ref{fig:vsini_radius_cdf}, we show the cumulative distribution functions (CDF) of \vsini\ for BSSs in the high- and low-density globular clusters and in the open clusters. Due to our \vsini\ measurement floor, values below our floor are set to 10 \kms; we use a permutation-based Kolmogorov-Smirnov test in which values below the floor are set to 10 \kms\ to model a null distribution built from the data for the following statistics. Two features stand-out. First, the CDFs of M67 and NGC 6791 are statistically significantly different ($p = 0.01$), with NGC 188 splitting the difference between the two, although not statistically significantly different than either ($p\simeq0.22$ for both comparisons). As clusters age, their MSTO and the temperatures of their BSS populations move toward lower temperatures below the Kraft break, increasing the fraction of BSSs that could spin-down. This naturally produces the observed sequence of open cluster CDFs. 

Second, as an integrated ensemble, the CDF of open clusters lies between the CDFs of low-density and high-density globular clusters, even though the open cluster central densities are an order of magnitude or more smaller than either globular cluster distribution. We find that the \vsini\ distribution of the three open clusters combined is different than those of both the high-density clusters ($p = 0.01$) and the low-density clusters ($p = 0.01$). These p-values should be conservative as we are not considering potential larger differences in the populations below the floor. 

Given that the open clusters have significantly lower densities than the globular clusters but their combined CDF lies between the lower- and higher-density globular clusters, there is a suggestion in this comparison that density may not be the only factor responsible for the fraction of fast rotators in a star cluster. 

\subsection{The Influence of Velocity Dispersion on BSS Rotation Distributions}\label{subsec:veloc_disp}

Here we suggest consideration of the role of velocity dispersion ($\sigma$) in determining BSS rotation distributions. In non-core collapse globular clusters, like in open clusters, binary mass transfer via Roche Lobe overflow has been suggested to be the primary channel to form BSSs, even in central regions \citep{leighWhereBlueStragglers2007,sollimaCorrelationBlueStraggler2008, ferraroBinaryrelatedOriginMediated2026}. Binary orbital period determines the evolutionary state when an evolving star will overflow its Roche lobe (for given companion mass). In a star cluster, the local velocity dispersion sets the hard-soft orbital period for binaries \citep[for example,][]{gellerINTERRUPTEDStelLARENCOUNTERS2015}. Thus the velocity-dispersion radial profile determines the maximum orbital periods present at different cluster radii, and thereby the most-evolved state of mass-transfer progenitors. Finally, because different evolutionary states may differ in stability and conservativeness of Roche lobe overflow (RLOF), mass transfer at different evolutionary states may produce BSSs of different masses and effective temperatures. If so, this would lead to different fractions of BSSs above and below the Kraft break, and thus different BSS rotation distributions. 

For specificity, we estimate the evolutionary states for mass transfer in two samples of non-core collapse globular clusters.
To begin, we use MIST stellar radii for giant stars in these globular clusters (initial masses of 0.8--0.9 \Msun\ and $-0.7<\rm[Fe/H]< -2$), the Eggleton approximation for the Roche lobe radius \citep[Equation 2 of][using secondary masses of mass ratio $q = \frac{M_{\rm accretor}}{M_{\rm donor}} = 1$ and $q = 0.6$]{eggletonAproximationsRadiiRoche1983}, and the hard-soft period boundary \citep[equation 1 of][using 0.3 \Msun\ as the average mass of cluster stars]{gellerINTERRUPTEDStelLARENCOUNTERS2015}. Main-sequence mass transfer may occur in systems with periods less than a few days, which would only be disrupted for $\sigma\gtrsim 50$ \kms. For various combinations of $q\;\rm{and\;[Fe/H]}$, binaries at the tip of the red giant branch (RGB) will have Roche lobe radii corresponding to orbital periods between 250 and 600 days for $18.0 \gtrsim \sigma\gtrsim11.0$ \kms. For thermally-pulsing asymptotic giant branch (AGB) stars at their maximum radii, orbital periods will range from 1300--1500 days for $9.5 \gtrsim \sigma\gtrsim7.6$ \kms. Finally, if wind RLOF (WRLOF) occurs in metal-poor binaries \citep{abateWindRochelobeOverflow2013}, WRLOF can occur in systems with periods of $10^4-10^5$ days \citep{sunWindRochelobeOverflow2024}, which correspond to $2.5 \gtrsim \sigma\gtrsim1.0$ \kms. 

The non-core collapse globular clusters of \cite{ferraroFastRotatingBlue2023} can be split into high central velocity dispersion ($\sigma_0\geq10$ \kms: 47 Tuc and NGC 1851; $\omega$ Centauri $\sigma_0\sim18$ \kms) and low central velocity dispersion ($\sigma_0\leq5$ \kms: M55, M4, and NGC 3201). The velocity dispersion profiles \citep{baumgardtCatalogueMassesStructural2018} of the high-$\sigma_0$ clusters indicate that main-sequence and RGB RLOF should dominate BSS production near the cores, whereas BSSs can be formed by RLOF from all evolutionary states in the envelopes of high-$\sigma_0$ clusters. Binaries from all evolutionary states also form by RLOF at all radii in the low-$\sigma_0$ clusters. WRLOF only occurs at the outermost radii of globular clusters. 

There are observational signs that velocity dispersion variation indeed may play a role in observed rotation distributions of globular cluster BSSs. For example, in $\omega$ Centauri the velocity dispersion profile \citep{sollimaNonpeculiarVelocityDispersion2009} only permits main-sequence and RGB mass transfer in the core but by 3--4 times the core, radius the velocity dispersion drops sufficiently that AGB mass transfer is also possible. This is the same region where \cite{ferraroFastRotatingBlue2023} (see their Figure 5) found a fast increase in the frequency of rapid rotators.

In Paper 1, we found that very-long-period progenitor binaries ($P>10^4$ days) must significantly contribute to the BSS population through AGB RLOF and WRLOF. Since these binaries are disrupted at most radii of globular clusters, the relative ratios of BSS formation across progenitor evolution states and even across mechanisms (e.g, mergers) could change their observed rotation distributions.  

This discussion is intended only to present the idea. There remain many unaddressed issues; for example, binaries move through regions of different velocity dispersions. Ultimately, detailed binary stellar evolution models of low-mass stars coupled with N-body simulations are needed to understand relative BSS production by formation mechanism and the consequences this has for final BSS mass, \Teff, and rotation period.

\section{Summary}\label{sec:summary}

In this paper, we find that rotation distributions of BSSs in old open clusters show a Kraft break very similar to that found in the field. Hotter BSSs have rotation periods under two days---still well below critical velocity---which we take to trace the degree of spin-up during the formation of all BSSs in these clusters. BSSs below the Kraft break exhibit very slow rotation periods, presumably due to magnetic braking linked to their convective envelopes. BSSs within the Kraft
break have a mixture of rotation rates. That these BSSs exhibit spin down only at the same temperatures at which field stars spin down indicates that at all effective temperatures they have similar structures in their envelopes even after having accreted mass from a companion. Because many known BSSs are in or hotter than the Kraft break, deriving gyrochronologic ages must be done with care. 

The fraction of rapidly rotating BSSs in open clusters is directly tied to the fraction of BSSs in and above the Kraft break. In metal-poor globular clusters, we found that stars with \Teff\ below $6896\pm 156$ K were significantly spun-down in comparison to the median rotation velocity above that temperature. This rotation transition is hotter by 100-250 K at low metallicities ([Fe/H] $\sim-1.5$) than at solar metallicity. However, above this temperature the globular cluster rotation distribution shows many more slowly rotating stars than found in or above the Kraft break in open clusters. 

Finally, we compare the \vsini\ distribution of BSSs in open and globular clusters and find that the density of formation environment is not the only possible cause for the fraction of fast and slow rotators in a cluster. For example, the local velocity dispersion at a region in a globular cluster can impact the available progenitor binaries and perhaps formation mechanisms for BSSs, which may also manifest in different \vsini\ distributions.

\begin{acknowledgements}
The authors express their gratitude to D. Dixon, E. Leiner, E. Motherway, R. S. Narayan, A. Nine, and R. Townsend for their insightful feedback during the creation of this manuscript and to the many undergraduate and graduate students of the R. D. Mathieu research group and the staff of WIYN observatory, without whom we would not have been able to collect thousands of stellar spectra that enabled the findings of this work. Finally, we acknowledge the support of the Wisconsin Alumni Research Fund and the Wisconsin Space Grant Consortium through awards RFP25\_4-0 and RFP25\_12-0.

This work has made use of data from the European Space Agency (ESA) mission Gaia (\url{https://www.cosmos.esa.int/gaia}), processed by the Gaia Data Processing and Analysis Consortium (DPAC; \url{https://www.cosmos.esa.int/web/gaia/dpac/ consortium}). Funding for the DPAC has been provided by national institutions, in particular the institutions participating in the Gaia Multilateral Agreement. 

This work was conducted at the University of Wisconsin-Madison, which is located on occupied ancestral land of the Ho-Chunk people, a place their nation has called Teejop since time immemorial. In an 1832 treaty, the Ho-Chunk were forced to cede this territory. The university was founded on and funded through this seized land; this legacy enabled the science presented here. Observations for this work were conducted at the WIYN telescope on Kitt Peak, which is part of the lands of the Tohono O’odham Nation.
\end{acknowledgements}

\facilities{WIYN - Wisconsin-Indiana-Yale-NOAO Telescope (Hydra MOS), Gaia}

\software{\textsf{Astropy} \citep{astropycollaborationAstropyCommunityPython2013,astropycollaborationAstropyProjectBuilding2018,astropycollaborationAstropyProjectSustaining2022}, \textsf{MIST} \citep{dotterMESAISOCHRONESLAR2016,choiMESAISOCHRONESSTELLAR2016,paxtonMODULESEXPERIMENTSSTELLAR2011,paxtonMODULESEXPERIMENTSSTELLAR2013,paxtonMODULESEXPERIMENTSSTELLAR2015}, \textsf{NumPy} \citep{harrisArrayProgrammingNumPy2020},  \textsf{SciPy} \citep{virtanenSciPyFundamentalAlgorithms2020}, \textsf{scikit-learn} \citep{scikit-learn}, \textsf{ChatGPT-4} \citep{openai_chatgpt_2025}}

\appendix
\section{BSS \vsini, \Teff, and Radius Measurements}\label{app:measurements}

Table~\ref{table:bss_vsini} contains the Gaia DR3 ID, WOCS ID, and Gaia DR3 RA and Dec \citep{GaiaDataRelease2023} of each BSSs in M67, NGC 188, and NGC 6791 along with \vsini\ measurements and the \Teff\ and radius found in Paper 1.

Several studies have spectroscopically measured temperatures of some of the BSSs in M67 and NGC 188. GALAH Survey DR4 \citep{buderGALAHSurveyData2025, kosGALAHSurveyImproving2025} has measured the $T_{\rm eff}$ of 13 of the M67 BSSs; the average absolute difference between their measurements and ours is 75 K with no systematic differences and a maximum difference of 130 K. \cite{nineWIYNOpenCluster2024} fit the temperatures of 15 of the M67 BSSs using the temperature-sensitive wing profiles of H$\alpha$; they found temperatures that were on average 220 K cooler than we found, with the largest deviations (600 K) among stars of $T_{\rm eff}>8000\;\rm K$ where the H$\alpha$ technique becomes less sensitive to temperature due to Stark broadening \citep{lovisiChemicalKinematicalProperties2012}. After removing any measurements flagged for bad values, APOGEE-2 has spectroscopic $T_{\rm eff}$ measurements of many of the hottest BSSs \citep[7 in M67 and 2 in NGC 188,][]{abdurroufSeventeenthDataRelease2022}. These measurements are on average $\sim$600 K cooler (range 370-1120 K) than MIST and GALAH DR4. Although these differences are significant, the lower temperatures do not change the temperature region of any BSSs in Section~\ref{subsec:teff_vsini}. 

We note that the relationship between Gaia colors and MIST-derived \Teff\ becomes steeper above 8000 K (Gaia BP$-$RP $<$ 0.2), so a small change in color leads to a larger change in temperature, which may explain some of the differences seen above. 

\begin{splitdeluxetable*}{cccccBccc} \label{table:bss_vsini}
% \tabletypesize{\footnotesize}
\tablecaption{BSS \vsini, \Teff, and Radius Measurements}
\tablehead{
   \colhead{Gaia DR3 ID} & \colhead{WOCS ID} & \colhead{Cluster} 
   & \colhead{RA (ICRS)} &  \colhead{Dec (ICRS)} 
   & \colhead{\vsini}  &\colhead{$T_{\rm{eff}}$}
  &\colhead{Radius} 
    \\
    \colhead{} & \colhead{} &\colhead{} & \colhead{} & \colhead{} & \colhead{\kms} &  \colhead{K} & \colhead{$\rm{R_\odot}$}}
\startdata
$573933322966008960$	& $8104$ & NGC 188	&$10.064616585556992	$&$85.06345774998789	$& $<10$&${6081} \,[{6039},\,{6127}]$	& ${1.11} \,[{1.10},\,{1.13}]$\\ 
$573937961530665088$	& $5467$& NGC 188	&$12.60415212043305	$&$85.18392229634908$	& $<10$& ${6126} \,[{6085},\,{6171}]$	& ${1.15} \,[{1.13},\,{1.16}]$\\ 
$573944111923749760$	& $4540$	& NGC 188	&$11.326406727485669$	&$85.32219238013796$ & 70 & ${6651} \,[{6613},\,{6693}]$	& ${2.29} \,[{2.27},\,{2.31}]$\\
\enddata
\tablecomments{This table is available in machine-readable format. A portion is shown here for guidance regarding its content. The first value is the average and the values in the square brackets are the value at the 16th and 84th percentiles, respectively. \vsini\ values below our floor are marked with $<10$ \kms.}
\end{splitdeluxetable*}

\bibliography{references}{}
\bibliographystyle{aasjournalv7}

\end{document}